# Novel Algorithm for Sparse Solutions to Linear Inverse Problems with Multiple Measurements

Lianlin Li, Fang Li

Institute of Electronics, Chinese Academy of Sciences, Beijing, China

*Lianlinli1980@gmail.com*

*ABSTRACT*: In this report, a novel efficient algorithm for recovery of jointly sparse signals (sparse matrix) from multiple incomplete measurements has been presented, in particular, the NESTA-based MMV optimization method. In a nutshell, the jointly sparse recovery is obviously superior to applying standard sparse reconstruction methods to each channel individually. Moreover several efforts have been made to improve the NESTA-based MMV algorithm, in particular, (1) the NESTA-based MMV algorithm for partially known support to greatly improve the convergence rate, (2) the detection of partial (or all) locations of unknown jointly sparse signals by using so-called MUSIC algorithm; (3) the iterative NESTA-based algorithm by combing hard thresholding technique to decrease the numbers of measurements. It has been shown that by using proposed approach one can recover the unknown sparse matrix $\bar{\bar{X}}$ with $Spark(\bar{\bar{A}})$-sparsity from $Spark(\bar{\bar{A}})$ measurements, predicted in Ref. [1], where the measurement matrix denoted by $\bar{\bar{A}}$ satisfies the so-called restricted isometry property (RIP). Under a very mild condition on the sparsity of $\bar{\bar{X}}$ and characteristics of the $\bar{\bar{A}}$, the iterative hard threshold (IHT)-based MMV method has been shown to be also a very good candidate.

I. INTRODUCTION

Recovery of sparse signals from a small number of measurements is a fundamental problem in many practical applications such as medical imaging, seismic exploration, communication, image denoising, analog-to-digital conversion, and so on. The well-known compressed sensing, developed by Candes, Tao and Donoho et al, studies information acquisition methods as well as efficient computational algorithms. By exploiting colorful results developed within the framework of compressive sensing, we can reconstruct a sparse vector $\bar{x}$ by solving the highly underdetermined linear equations $\bar{y} = \bar{\bar{A}}\bar{x}$ under minimal $\ell_1$-norm constraint only if the measurement matrix $\bar{\bar{A}}$ satisfies some properties such as restricted isometry property (RIP), null-space property (NSP), and so on. Though determining the sparest vector $\bar{x}$ consistent with the data $\bar{y} = \bar{\bar{A}}\bar{x}$ is generally an NP-hard problem, many suboptimal algorithms have been formulated to attack this problem, for example, greedy algorithm, basis pursuit (BP), Bayesian algorithm, and so on.

The single measurement sparse solution problem has been extensively studied in the past. In many practical applications such as dynamic medical imaging, neromagnetic inverse problem, beam forming, electromagnetic inverse source, communication, and so on, the recovery of jointly sparse signal or MMV problem, the variation of the compressive sensing or sparse linear inverse problem, is an important topic; in particular, to deal with the computation of sparse solution when there are multiple measurement vectors (MMV) and the solutions are assumed to have a common sparsity profile or jointly sparse. The most widely studied approaches to the MMV problem are based on solving the convex optimization problem

$$\min_{\bar{\bar{X}}} \left\| \bar{\bar{X}} \right\|_{p,q}, \text{ subject to } \bar{\bar{B}}_{n \times L} = \bar{\bar{A}}_{n \times N} \bar{\bar{X}}_{N \times L} \tag{1}$$

where the mixed $\ell_{p,q}$ norm of $\bar{\bar{X}}$ is defined as

$$\left\| \bar{\bar{X}} \right\|_{p,q} = \left\| \tilde{X} \right\|_p$$

with $\tilde{X} = \left[ \left\| \bar{\bar{X}}(1,:) \right\|_q, \left\| \bar{\bar{X}}(2,:) \right\|_q, \cdots, \left\| \bar{\bar{X}}(N,:) \right\|_q \right]^T$, $\bar{\bar{X}}(j,:)$ is the $j$th row of $\bar{\bar{X}}$.

Up to now, many efforts have made to attack this problem. Cotter et al. considered the minimization problem of [2]

$$\min_{\bar{\bar{X}}} \left( \left\| \bar{\bar{X}} \right\|_{p,2} \right)^p \text{ subject to } \bar{\bar{B}}_{n \times L} = \bar{\bar{A}}_{n \times N} \bar{\bar{X}}_{N \times L}.$$

Chen and Huo considered the uniqueness under $p = 0$ via the spark of measurement matrix $\bar{\bar{A}}$ and equivalence between the minimization problem with $p = 1$ and $p = 0$ [1]. Further, the orthogonal matching pursuit (OMP) algorithm for MMV has also been developed [1]. Tropp dealt with (1) for $p = 2$ and $q = \infty$. Mishali and Eldar proposed the ReMBo algorithm which reduces MMV to a series of SMV problems. Eldar and Rauhut proposed the OMP algorithm with hard threshold technique and analyzed the average case for jointly sparse signal recovery [4]. Berg and Friedlander studied performance of $\ell_{1,1}$ and $\ell_{1,2}$ for different structure of sparse $\bar{\bar{X}}$ [5].

In this presentation, we consider in depth the extension of a class of algorithm—NESTA algorithm—to the multiple measurement vectors available, and solutions with a common sparsity structure must be computed, especially, NESTA-based MMV algorithm. Inspired by recent breakthroughs in the development of novel first-order methods in convex optimization, the cost functions appropriate to NESTA-based MMV are developed, and algorithms are derived based on their minimization. Further several approaches to improve the NESTA-based MMV algorithm to decrease the number of measurements and increase the convergence rate have been proposed. This report demonstrates that this approach is ideally suited for solving large-scale MMV reconstruction problems.

## II. ALGORITHMS

In this section, the basic idea of NESTA-based MMV algorithm has been provided; moreover, several approaches to improve it have been discussed. We will refer the reader to [7] for detailed discussions about proposed approaches.

### II.1 NESTA-based MMV Algorithm

Similar done by Nesterov's method for single-measurement problem [9] [10], the NESAT-based MMV algorithm minimize the smooth convex function $f$ on the convex set $Q_p$,

$$\min_{\bar{\bar{X}} \in Q_p} f(\bar{\bar{X}}) \tag{2}$$

where the primal feasible set $Q_p$ is defined by

$$\bar{\bar{X}} \in Q_p := \left\{ x : \left\| \bar{\bar{A}}\bar{\bar{X}} - \bar{\bar{B}} \right\| \leq \varepsilon \right\}.$$

To exploit fully the structure of unknown matrix $\bar{\bar{X}}$, we introduce $\bar{\bar{X}} = \bar{\bar{\Psi}}\bar{\bar{\alpha}}$ with sparse transformation matrix $\bar{\bar{\Psi}}$. The smoothed version of convex function $f(\bar{\bar{X}})$ in (2) is

$$f_\mu(\bar{\bar{\alpha}}) = \max_{\bar{\bar{U}} \in Q_d} \left\{ \left\langle \bar{\bar{U}}, \bar{\bar{\alpha}} \right\rangle - \mu p_d(\bar{\bar{U}}) \right\} \tag{3}$$

where $Q_d := \left\{ \bar{u} : \left\| \bar{\bar{U}} \right\|_\infty \leq 1 \right\}$ is the dual feasible set.

To control flexibly the inherent structure of $\bar{\bar{X}}$, the smoothed convex function (3) is proposed to be rewritten as

$$f_\mu(\bar{\bar{\alpha}}) = \max_{\bar{u} \in Q_d} \left\{ \left\langle \bar{u}, \tilde{\bar{\alpha}} \right\rangle - \mu p_d(\bar{u}) \right\} \tag{4}$$

where

$$\tilde{\bar{\alpha}} = \left[ m(\bar{\bar{\alpha}}(1,:)), m(\bar{\bar{\alpha}}(2,:)), \cdots, m(\bar{\bar{\alpha}}(N,:)) \right]^T,$$

$m(\bar{\bar{\alpha}}(j,:))$ is a homogeneous function of $j$th row of $\bar{\bar{\alpha}}(j,:)$,

$p_d(\bar{u})$ is a prox-function for dual feasible set $Q_d$ denoted by $Q_d := \left\{ \bar{u} : \left\| \bar{u} \right\|_\infty \leq 1 \right\}$.

As done by standard NESTA's method for recovery of single-measurement sparse signal, one has the procedure of NESTA-based MMV algorithm shown in Table 1

From Table 1, it is noted that (1) $\nabla f_\mu(\bar{\bar{\alpha}})$ can be easily computed in the closed form, (2) the proposed algorithm belongs to the first-order method for constraint optimization problem,(3) if the row of $\bar{\bar{A}}$ is orthogonal, which is often the case in compressed sensing applications [9], the computational cost is very low, in particular, each iteration is extremely fast. To decrease the number of measurements and increase the convergence rate, the following approaches are carried:

(1) As done in [9], the homotopy technique can be exploited to accelerate it.

(2) It has been empirically shown if partial support of unknown sparse matrix $\bar{\bar{\alpha}}$, the convergent speed will be improved; moreover, the number of measurements can be greatly decreased. If the partial common locations of $\bar{\bar{\alpha}}$ denoted by $T$ are known, the function (3) is modified as

$$f_\mu(\bar{\bar{\alpha}}) = \max_{\bar{\bar{U}} \in Q_d} \left\{ \langle \bar{\bar{U}}, \bar{\bar{\alpha}}_{T^c} \rangle - \mu p_d(\bar{\bar{U}}) \right\} \qquad (3m)$$

Of course, (4) can be modified as

$$f_\mu(\bar{\bar{\alpha}}) = \max_{\bar{u} \in Q_d} \left\{ \langle \bar{u}, \tilde{\bar{\alpha}}_{T^c} \rangle - \mu p_d(\bar{u}) \right\} \qquad (4m)$$

Of course, the size of $\bar{u}$ in (4m) will be smaller than one in (4).

(3) To estimate the partial support of jointly sparse matrix $\bar{\bar{\alpha}}$, the so-called MUSIC algorithm can be exploited. As we known, if the more column-rank of $\bar{\bar{\alpha}}$ is, the more support of $\bar{\bar{\alpha}}$ can be obtained.

(4) To decrease the number of measurements, the iterative NESTA-based MMV algorithm by combing hard threshold technique is carried out, see Table 2,

TABLE1. The procedure of NESTA-based MMV algorithm

Initialize $\bar{\bar{\alpha}}_0$. For $k \geq 0$

Step1. *Compute* $\nabla f(\bar{\bar{\alpha}}_k)$

Step2. *Compute* $\bar{\bar{y}}_k$:

$$\bar{\bar{y}}_k = \arg\min_{\bar{\bar{\alpha}} \in Q_p} \left\{ \frac{L}{2} \|\bar{\bar{\alpha}} - \bar{\bar{\alpha}}_k\|_2^2 + \langle \nabla f(\bar{\bar{\alpha}}_k), \bar{\bar{\alpha}} - \bar{\bar{\alpha}}_k \rangle \right\}$$

Step3. *Compute* $\bar{\bar{z}}_k$:

$$\bar{\bar{z}}_k = \arg\min_{\bar{\bar{\alpha}} \in Q_p} \left\{ \frac{L}{\sigma_p} p_p(\bar{\bar{\alpha}}) + \sum_{i=0}^{k} \alpha_i \langle \nabla f(\bar{\bar{\alpha}}_i), \bar{x} - \bar{\bar{\alpha}}_i \rangle \right\}$$

$$\alpha_i = \frac{i+1}{2}$$

Step4. *Update* $\bar{\bar{\alpha}}_k$:

$$\bar{\bar{\alpha}}_k = \tau_k \bar{\bar{z}}_k + (1 - \tau_k) \bar{\bar{y}}_k$$

$$\tau_k = \frac{2}{k+3}$$

Stop *when a given criterion is valid*

Table 2. Iterative NESTA-based MMV algorithm

```
Initialize  $\bar{\bar{\alpha}}$.
Step 1.  Carrying NESTA – based MMV algorithm with partially
         known suppor provided in Table 1
Step 2   If a given criterion is valid;
            stop
         esle
            Choose support T shown in Eq.3(m) or Eq.4(m) via hard threshold,
            GoTo   STEP 1
         end
```

II.2  IHT-based Algorithm

As a matter of fact, the iterative hard thresholding algorithm proposed by Blumensath and Davies can be easily generalized to deal with MMV problem (see Table 3). Further, the theoretical analysis about performance guarantee can be carried out along the same line as done in [8]

Table 3. Iterative hard threshold based MMV algorithm

```
Initialize   $\bar{\bar{\alpha}} = \bar{\bar{\Phi}}^T \bar{y}, where, \bar{\bar{\Phi}} = \bar{\bar{A}}\bar{\bar{\Psi}}$
DO
    $\bar{\bar{R}}^n = \bar{\bar{\alpha}}_K - \bar{\bar{\alpha}}^n$
    $\bar{\bar{a}}^{n+1} = \bar{\bar{\alpha}}^n + \mu\bar{\bar{\Phi}}^T \left( \bar{y} - \bar{\bar{\Phi}}\bar{x}^n \right)$
    $\bar{\bar{\alpha}}^{n+1} = H_K \left( \bar{\bar{a}}^{n+1} \right)$
Stop   a given criterion is valid
```

III.  CONCLUSIONS

In this report, we focus on the NESTA-based algorithm to deal with the recovery of jointly sparse signal. Numerical experiences tell us that the NESTA-based MMV algorithm can be used to deal with the large-scale MMV problem. Moreover, the proposed approach outperforms the stat-of-art algorithms for MMV problem. Detailed discussion about our algorithms, with necessary theoretical analysis will appear in [7].